\documentclass[submission,copyright,creativecommons]{eptcs}
\providecommand{\event}{SCSS 2021} 
\usepackage{breakurl}             
\usepackage{underscore}           

\title{Symbolic Computation in Software Science:\\
	My Personal View}
\author{Bruno Buchberger
\institute{Research Institute for Symbolic Computation (RISC)\\
	Johannes Kepler University\\
	Linz / Schloss Hagenberg, Austria}
\email{Bruno.Buchberger@risc.jku.at}
}
\def\titlerunning{Symbolic Computation in Software Science: My Personal View}
\def\authorrunning{B. Buchberger}
\begin{document}
\maketitle

\begin{abstract}
In this note, I develop my personal view on the scope and relevance of symbolic computation in software science. For this, I discuss the interaction and differences between symbolic computation, software science, automatic programming, mathematical knowledge management, artificial intelligence, algorithmic intelligence, numerical computation, and machine learning. In the discussion of these notions, I allow myself to refer also to papers (1982, 1985, 2001, 2003, 2013) of mine in which I expressed my views on these areas at early stages of some of these fields.
\end{abstract}

\section*{The Intention of This Note}

It is a great joy to see that the SCSS (Symbolic Computation in Software Science) conference series, this year, experiences its 9th edition. A
big Thank You to the organizers, referees, and contributors who kept the series going over the years! The series emerged from a couple of meetings
of research groups in Austria, Japan, and Tunisia, including my Theorema Group at RISC, see the home pages of the SCSS series since 2006. In 2012,
we decided to define ``Symbolic Computation in Software Science'' as the scope for our meetings and to establish them as an open conference series
with this title.

As always, when one puts two terms like ``symbolic computation'' and ``software science'' together, one is tempted to read the preposition
in between - in our case ``in'' - as just a set-theoretic union. Pragmatically, this is reasonable if one does not want to embark on scrutinizing
discussions. However, since I was one of the initiators of the SCSS series, let me take the opportunity to explain the intention behind SC\textit{
	in }SS in this note. Also, this note, for me, is a kind of revision and summary of thoughts I had over the years on the subject of SCSS and related
subjects. Hence, allow me to refer to a couple of my papers with basic considerations on the subject. I do not discuss, however, any of my technical
contributions to the subject of SCSS (which would be, mainly, Gr{\"o}bner bases and the Theorema system). In some way, this note  continues, updates,
and specializes the note on mathematics in the 21st century  I gave at the beginning of SCSS 2013, see~\cite{scss2013}, from which I quote:

\begin{quote}
In my view, mathematics of the 21st century will evolve as a unified body of mathematical logic, abstract structural mathematics, and
computer mathematics with no boundaries between the three aspects. Working mathematicians will have to master the three aspects equally well and
integrate them into their daily work. More specifically, working in mathematics will proceed on the object level of developing new mathematical content
(abstract knowledge and computational methods) and, at the same time, on the meta-level of developing automated reasoning methods for supporting
research on the object level. This massage of the mathematical brain by jumping back and forth between the object and the meta-level will guide mathematics
onto a new level of sophistication. Symbolic computation is just a way of expressing this general view of mathematics of the 21 st century and it
also should be clear that software science is just another way of expressing the algorithmic aspect of this view.
\end{quote}

Continuing the discussion on the intended meaning of ``Symbolic Computation in Software Science'' in this note will hopefully help to advocate
the central importance of this topic for the future of mathematics, logic, and computer science. This should also motivate more and more people to
submit papers to the conferences in the SCSS series.

\section*{What is Symbolic Computation?}

In 1984, Academic Press London issued a call for designing a new journal for a new field that had emerged approximately since 1960. Various names
were in use for this field: computer algebra, symbolic and algebraic manipulation, analytic computation, formula manipulation, computation in finite
terms, symbolic computation, and others.  As a response to this call, I submitted a proposal to Academic Press for a ``Journal of Symbolic Computation''.
My proposal was selected and my clarification of the scope of ``symbolic computation'' formed also the Editorial of the journal, see \cite{jsc1985editorial}. 

I defined ``\textit{symbolic computation}'' as the area that deals with algorithms on symbolic objects, and I proposed ``\textit{symbolic
	objects}'' to be defined as finitary representations of infinite mathematical entities. Here, ``\textit{finitary}'' means ``storable in
a computer memory''. For example, finitely many generators with finitely many relations between words formed from the generators form a finitary
object that may represent an infinite group  (or, at least, a ``large'' group, i.e. a group whose number of elements is much much larger than
the size of the finitary representation). Algorithms can only work on finitary objects and the flavor of ``symbolic'' is exactly the point that
we want to solve problems on infinite (or ``large'') mathematical entities by finding algorithms that work on finitary (small), ``symbolic'',
representations of these entities. Also, numerical computation works on finitary representations (for example, lists of rational numbers that represent
a function consisting of infinitely many pairs of real numbers). In this sense, numerical computation is a subfield of symbolic computation. However,
usually, for algorithms to be called ``symbolic'' we request that the representation of the abstract mathematical domains by finitary domains
must be an isomorphism w.r.t. to the operations on the objects we study. In numerical computation, for the sake of efficiency, this request has to
be given up. Instead, we are satisfied with ``approximations''.

Pragmatically, in the editorial of the Journal of Symbolic Computation, I named three main areas for symbolic computation: \textit{computer algebra,
automated reasoning, and ``automatic programming''}. I also emphasized that all aspects of these areas should be in the scope of the Journal
of Symbolic Computation: mathematical theory on which symbolic algorithms can be based, the algorithms with their correctness proofs and complexity
analysis, the details of the implementation of the algorithms,  languages and software systems for symbolic computation, and applications. Indeed,
the three main branches of symbolic computation consider three important classes of ``symbolic objects'':

\begin{itemize}
\item[--] \textit{computer algebra}: symbolic objects that represent algebraic entities like terms that represent functions, differential operators, etc.
or finite relations that represent residue class structures;
\item[--] \textit{automated reasoning}: symbolic objects containing (quantified) variables that are considered as statements on (infinite) domains;
\item[--] \textit{automatic programming}: symbolic objects containing variables that are considered as programs that define processes on potentially infinitely
many inputs.
\end{itemize}

(Of course, these three sub-areas of symbolic computation are intimately connected and, in some precise way, even embedded in each other. The distinction
between the three areas is more or less only a matter of ``flavor''.)

In other words, symbolic objects are finitary objects that have ``semantics'' attached to them where, typically, the semantics is ``large'',
 even infinite, not tangible by computers whereas the symbolic objects are ``small'', finitary, tangible by algorithms. Any field of mathematics
can be studied under the ``symbolic'' view and,  actually, in any field of mathematics, if we want to solve problems by algorithms, we have
to find finitary representations for the objects in the field. Finding suitable finitary representations, by itself, may be a difficult - sometimes
provably impossible - mathematical problem: Before embarking on deeper questions, deciding whether or not two symbolic objects represent the same
abstract mathematical object and finding ``canonical'' representatives for symbolic objects may already be very difficult (sometimes provably
impossible). By finding representations of mathematical objects in any field of mathematics, the field becomes ``algebraic'', and problems in
the algebraic disguise of the field, essentially, become combinatorial problems. Thus, very sketchy, one may say: \textit{symbolic computation,
	ultimately, is the ``combinatorization'' of all of mathematics via finitary representations of infinite mathematical entities.}

It is a common misunderstanding that symbolic computation is the trivial side of mathematics, i.e. some people believe that, whereas ``pure''
mathematics lives in difficult spaces needing deep and difficult thinking, algorithmic mathematics (which must be ``symbolic'' in the above sense)
``just'' puts everything to the computer and presses the start button. The truth is, that the ``just'' needs more and deeper mathematics
than a mathematics that allows non-algorithmic constructions for problem-solving like the unlimited set quantifier, infinite summation, infinite
unions, transition to residua class domains etc. (A trivial example: In ``pure'' mathematics, a Gr{\"o}bner basis for given ideal generators
can be ``easily'' found by just taking the ideal generated by the generators. However, the definition of the ideal generated by generators involves
an infinite set construction!) Hence, with some provocation, in my view, mathematics only \textit{starts} at the moment when it tries to solve problems
by ``symbolic computation''. 

Recently, in 2020, we issued a call for running for the editor-in-chief position of the Journal of Symbolic Computation (JSC). At that occasion,
we asked the candidates to submit also their views on the scope of ``symbolic computation'' and of the JSC. Interestingly, the view of symbolic
 computation in the editorial of the JSC (and summarized above) was backed by all candidates and, basically, no dramatic changes or extensions
were proposed except that ``artificial intelligence'' was mentioned a couple of times.

Mentioning artificial intelligence, for me, raises some nostalgia because, when I founded the Research Institute for Symbolic Computation (also in
1985), for some time I was torn between using ``symbolic computation'' or ``artificial intelligence'' as the main notion in the name of the
new institute. At that time, bringing symbolic computation under the umbrella of artificial intelligence was quite tempting and also quite common:
For example, finding symbolic integrals was considered an ``artificial intelligence'' task like playing chess, with lots of heuristics. Correspondingly,
the most comprehensive symbolic computation software system at that time, MACSYMA, had ``MAC'' ($=$ Machine Aided Cognition) in its name! And,
of course, implementing heuristics is still a very important approach for improving the practical efficiency of methods for symbolic computation
problems. However, in 1985, I deliberately decided against having ``artificial intelligence'' in the name of my institute since I wanted to emphasize
the logical, mathematical, formal approach to problem-solving over the psychological, experimental aspect, which some people (then and now) believe
that goes ``beyond mathematics''.  I will go deeper into the analysis of the relationship between symbolic computation and artificial intelligence
later in this note.

Anyway, although symbolic computation (in the sense of the editorial of the JSC) seems to be a quite established and  stable notion, as a matter
of fact, in the JSC over the years one can observe that

\begin{itemize}
\item[--] the majority of papers in the JSC is on computer algebra,
\item[--] more and more, but still much fewer, papers are in automated reasoning,
\item[--] only a few papers came in on automatic programming.
\end{itemize}

\section*{Symbolic Computation in Software Science}

\textit{Software science} is the science of the process of developing software. This process starts from problems in some ``reality'' (part
of the real world) and creates software that solves the problems in an appropriate finitary model of this reality. Since the beginning of the software
age, the software development process has matured from being a kind of ``magic'' and being an ``art'' to a decent engineering discipline
called ``software engineering''. In parallel, since the very beginning, people have also tried to establish a ``science'' of software and
the software development process to make the process more reliable, provably correct, faster, more flexible, more economic, and ultimately automatic
or semi-automatic. Research in this direction is mostly summarized under the heading ``theoretical computer science''. Interestingly, the term
``software science'', which seems quite natural to me, in comparison to ``theoretical computer science'', is only used quite rarely. (This
can easily be verified by googling the two notions and comparing the number of relevant results.) 

Anyway, I think that ``software science'' is a quite useful notion that focuses on the actual development process of software and on its automation
and, hence, has a high impact on one of the central technologies - if not the central technology - of our age.

Since the objects of software science are formal models (domains with finitary objects and algorithmic operators on the objects), automation or
semi-automation of the software design process is essentially a ``symbolic computation'' process according to the definition of symbolic computation
we considered above. In other words, it should be clear that symbolic computation is the area that naturally should include also the (semi-)automation
of the software development process. Unfortunately, this logical analysis did not really create a big stream of papers on automating software development
to the JSC (and neither to conferences in the area of symbolic computation like ISSAC, ACA, SYNASC, etc.). Therefore, in 2012, I argued that the topic
``\textit{Symbolic Computation in Software Science}'' could and should get special attention by turning our group meetings into a conference
series with this name. 

Still, the idea that symbolic computation should have a major application in software science - in particular in the automation of the software development
process - did not create a big echo in the symbolic computation community. Neither do many people who work in software engineering realize that the
automation of the software development process is essentially a symbolic computation task. One reason for this is, surely, that there are strong
conference series and journals in the area of automated reasoning and related subjects.  (A side-remark: As some readers may know, when I built
up RISC starting from 1985, I also devoted much of my time to building up the ``Softwarepark Hagenberg''. With this, I wanted to demonstrate
that the mathematically deep field of symbolic computation has also the power to create something with a strong practical impact:  I started the
Softwarepark with 25 people.  When I stepped back as the director of the Softwarepark Hagenberg in 2013, 2500 people working and/or studying in
the Softwarepark.  I hoped that the government and the administration of my university would have noticed and recognized the unique power that
RISC / symbolic computation had created and had turned into innovation in the software foundations, into software development, and into software
business. Therefore, in 2013, I asked the government and university administration to establish an extra professorship ``Software Science'' in
the frame of RISC with the task of continuing my work for directing and expanding the Softwarepark based on solid research on symbolic computation
in software science. Indeed, in response to my request and argumentation, a professor position for ``Software Science'' was created but then,
much to my displeasure, giving in to the pressure of the informatics department, the position was finally used for something ``more useful''
for the education of the informatics undergraduates.)

Now, what I called ``automatic programming'' in the preface of the JSC, could also be called ``symbolic computation in software science''.
In more detail, I want to make this clear in this note. If seen in the right way, I think that symbolic computation in software science is / could
be / should be the most / one of the most fascinating topics of the next stage of mathematics / logic / computer science. (I like to call mathematics,
logic, and computer science together  ``thinking technologies'' or just ``full-stack mathematics''. Unfortunately, ``mathematics''
sounds old-fashioned to some people, sounds ``non-creative'' to others, boring to others, intimidating to again others. However, one may bend
and turn this as one likes, finally, at the top of the creative hierarchy of problem-solving and gaining knowledge by thinking, there is mathematics
at higher and higher levels - whether certain people in politics, science, economy, philosophy, culture, media or the people at the beer table like
it or not.)

\section*{A Stream of Problems on the Way}

On the way from a problem description / a collection of problem descriptions to an algorithm / program / software system that solves the problem
there are many creative steps each of which can be handled ad hoc for the particular problem at hand by a mathematician, computer scientist, developer.
Each of these steps, however, can also be considered as a problem on the meta-level with some symbolic objects (like software requirements, programs,
algorithm schemata, verification conditions, etc.) as input and symbolic objects as output for which we would like to have a general algorithmic
solution.

In this section, we assume that all the symbolic objects on the way from a problem specification/requirements to an algorithm  / piece of software
are expressions that describe or at least try to describe something ``in general terms''. In particular, we assume that the problem specification
(even a vague attempt of a specification that may need much clarification and reformulation) tries to describe the problem in general and not only
by examples. 

The important case that a problem specification, for certain reasons, can only be given by examples and cannot be explained in general terms, is
analyzed in detail in the next section in the paragraph on machine learning.

In the majority of cases, problem specifications are explicit in the sense that they are specified by an expression $P[x, y]$ with input
variable(s) $x$ and output variables(s) $y$, and a solution algorithm $A$ has to satisfy $P[x, A[x]]$, for all $x$. (However, there are
important classes of algorithmic problems that cannot be described in explicit form. For example, a canonical simplifier $A$ for an equivalence
relation $P$ cannot be described in this form. More generally, for example, the specification of operations on data structures by axioms
or the construction of algorithmic isomorphic representations of mathematical domains is not an explicit specification. We cannot go into more details
about this here.) 

Depending on the situation, the initial (often vague) problem descriptions may be given in natural language, maybe mixed with drawings and diagrams,
or in some formal language. 

Also, it is important to distinguish between two extremes:

\begin{itemize}
	\item Finding algorithms for fundamental, non-trivial, \textit{stand-alone algorithmic problems}: In this case, the problem specification and the solution
	algorithm are completely formal, symbolic objects and everything that happens between problem and solution should be amenable to algorithmic treatment
	on the meta-level, i.e. to symbolic computation. For such problems, typically, time and memory complexity is an issue. Examples: the problem of finding
	shortest paths in graphs; the problem of finding symbolic integrals; the problem of finding Gr{\"o}bner bases; etc.
	\item Developing software for \textit{an entire application}: In this case, the individual parts of the system (called ``units'') should implement
	a (big) number of functionalities, most of which are not really difficult. Only some of the functionalities may involve the algorithmic solution
	of fundamental problems. The algorithms for these functionalities, typically, are known and taken from reliable sources. The complexity of such systems
	originates from the huge number of units and the various (desired and undesired) interactions of the units. Also tuning of the known algorithms to
	the application at hand is an issue.
\end{itemize}

This distinction is important for the following reason: The application of formal methods for establishing the correctness of software only makes
sense if we consider non-trivial algorithmic problems. In contrast, for most of the millions of units to be developed in large software systems a
formal specification of the problem to be solved by the unit would be essentially identical to the code to be developed. In other words, a proposal
for the code of a unit, in the case of ``easy'' problems, is a way for describing the problem to be solved. This is the reason why rapid prototyping
and agile software development, in such situations, is so useful. It is also the reason why formal algorithm verification methods are rarely used
in the practice of developing large software systems. \vspace{0.3cm}

\noindent\textbf{Example.} In a calendar software system, probably, we want one unit that should check whether a proposed new calendar entry collides with
one of the existing entries. Let us assume that a calendar entry is characterized by its start time and end time.  The input to the unit will then
consist of four time moments $ s_1, e_1, s_2, e_2$  for the start time and the end time of the first and the second entry, respectively, with input
condition $ s_1< e_1$  and  $ s_2< e_2$. ``After some thinking'', the problem will then be described by most developers by a sentence like
this: ``The two calendar entries characterized by  $ s_1, e_1, s_2, e_2$   collide iff   $ s_2 \leq s_1 \leq  e_2$ or
$ s_1\leq s_2 \leq e_1$.'' Now it is clear that this ``specification'' of the problem is, basically, already the solution algorithm.
Only some transformation into the syntax of the programming language used is necessary. No powerful algorithm verification method or algorithm synthesis
method is necessary in such a case.  \vspace{0.3cm}

As simple as the example is, it is not too simple to guarantee the avoidance of severe flaws in the development. I tested the example out by presenting
it to various (reasonably experienced) developers. Amazingly, a few came up with the following specification / code: ``The two calendar entries
characterized by  $ s_1, e_1, s_2, e_2$  collide iff either $ s_2 \leq s_1 \leq  e_2$ or $ s_2\leq e_1 \leq e_2$.''
This specification is ``incorrect'' because it does not include the case $ s_1<s_2<e_2<e_1$, which of course ``everybody'' would also consider
as a collision, even a ``particularly heavy one''. (I put ``incorrect'' in quotation marks because, at the very first stage of uttering a
request, the ``customer is always right''. Maybe, he really wants what he tells! Either one just implements what he tells or one may consider
the subsequent discussion as a way to find out what he ``really wants'' or to change his mind about what he wants.) This shows that already in
the ``thinking'' between a vague indication of a problem and its specification (by a general statement, not only by examples), severe mistakes
may be made (or, considered differently, the request of the customer may undergo serious changes). In our example, we also could start a little ``earlier''
and just say: ``The two calendar entries characterized by  $ s_1, e_1, s_2, e_2$   collide iff  the time interval $ \left[s_1,e_1\right]$ intersects
with the time interval $ \left[s_2,e_2\right]$.'' Now we could question the notion ``intersects'' and might agree on the following:  ``The
two calendar entries characterized by  $ s_1, e_1, s_2, e_2$   collide iff  there is a time moment \textit{x} such that  $ s_1\leq x\leq
e_1$  and $ s_2\leq x\leq e_2$.'' In this form, we can send the condition into a quantifier elimination algorithm and we will get an answer which
will be equivalent to ``$ s_2 \leq s_1 \leq  e_2$ or $ s_1\leq s_2 \leq e_1$.'' (Please try it out, it is worthwhile!)
Hence, this simple example shows that, actually, already in the very early stage of discussing and clarifying even seemingly simple requirements
a lot of systematic/formal thinking is involved, which in principle should be amenable to automating and, hence, symbolic computation!

Thus, we start at the very early stage of having a vague desire of achieving something by software and go through all the stages of developing a
piece of software that fulfills the desire and, further, through all the stages of maintaining, updating, improving, and integrating pieces of software
to fulfill more and more sophisticated desires. Through all these stages, we ask ourselves how much of this process can be (semi-)automated. This
gives a rich list of R$\&$D topics, which make up \textit{the important topics in the scope of ``symbolic computation in software science''
	as described in the calls for the SCSS series}, see the latest version in the call for SCSS 2021.  This call contains, roughly, 20 important and
quite diverse but strongly interconnected topics on the way from requirements to software.

I do not list these topics and comment on all these topics here. Rather, let me give some personal remarks that emphasize, and maybe expand, some
of the subjects, themes, and objectives of the topics in the SCSS calls.

\begin{itemize}
\item My feeling is that relatively little research is available on (semi-)automating the development of \textit{large software systems consisting of
tons of simple ``units'' }of the type we have seen above in the example. Research has focused more on symbolic methods for algorithm verification
and synthesis for non-trivial algorithms. In some way, this is unfortunate because the construction of tons of software is necessary today, semi-automation
of this process is needed and could be a big business. Our research results are too much oriented on automating the invention of ``important'',
``difficult'' algorithms.  However, the (semi-\nolinebreak[4]) automation of the development of huge amounts of simple programs and their interaction, in some
way, is quite challenging, much needed, and asks for formal methods to guarantee the quality of the process.

\item As a variant of developing big software systems consisting of many simple units we also should consider the task of re-engineering big software systems
that were written years ago in programming languages that are antiquated now. Often, the documentation of such systems is lost or fragmentary, and
finding out what the units should do, i.e. getting a problem specification from code, is a major task.

\item In most cases, software development starts from vague requirements in \textit{natural language (maybe with diagrams or drawings).} The task is to
come up, maybe in an interactive dialogue, with a bunch of formal requirements. Here, we should allow natural language or, maybe, a simplified version
of natural language as a symbolic language: The sentences that formulate requirements are ``finitary'' with infinite semantics since, normally,
requirements have hidden universal quantifiers in it. (See the simple example above: The requirement is formulated for arbitrary calendar entries.
In our first step towards formalization, the hidden universal quantifier goes over $ s_1, e_1, s_2, e_2$.) We also should allow diagrams or drawings
as symbolic objects: They are surely ``finitary'' and, usually, have infinite semantics, since a drawing normally tries to convey the important
features of infinitely many possible individual situations. Specifying requirements by natural language text or drawings is very different from the
specification of requirements by finitely many input/output pairs, see the analysis of ``machine learning'' in the next section. \textit{Allowing natural language or drawings in requirements} is of course a big challenge but I think we should take this deliberately under the umbrella of SCSS
because it will need much more than just ordinary NLP (natural language processing) and graphics. Rather, a systematic connection to logic is necessary.
(In fact, in dynamic geometry systems a lot of work in this direction is already done when graphical input explaining geometrical situations ``in
general positions'' is allowed.)

\item As we have seen in the simple example above, we also would like to go a step further and go from requirements in natural language and/or drawings
right away to algorithms/programs that satisfy the requirements. As we have argued in the example,  in the majority of ``units'' in application
software systems this will not be significantly more difficult than coming up with formal requirements.

\item The individual algorithms/programs in software systems do not live in empty air but inside a whole hierarchy of data structures and domains which,
depending on the context, are called (algorithmic) ``models'' of reality. Such models consist of problem specifications, definitions of notions,
knowledge, algorithms, and - in the ideal case - arguments/proofs why the operations/algorithms in the system meet their specifications. Hence, seen
in this way, software systems can also be considered mathematical knowledge systems. Hence, (semi-)automation of building and maintaining such systems
can also be seen under the umbrella of \textit{Mathematical Knowledge Management (MKM). }We introduced this term a couple of years ago in a slightly
different context, see the preface of the proceedings~\cite{mkm01}, which were expanded as the special issue~\cite{DBLP:journals/amai/BuchbergerGH03}. We propose that SCSS and MKM should be considered
together and, maybe, SCSS and MKM should be collocated in the future.

\item In practice, the correctness of software is established by testing rather than by formal verification. \textit{Testing }is a highly developed ``technology''
in software engineering: There is an arsenal of ``automated software testing'' systems available. They are well integrated into the various software
development environments and they are quite helpful for managing large test suites for the consecutive versions of software systems. However, I think
that much more could be done by applying formal methods for coming up with complete systems of test data from a given problem specification and program.
Here, completeness means that we would get one test input/output for each equivalence class of inputs that generate the same program path during
execution. Of course, in general, the set of these equivalence classes is not finite. However, in the practical case of large software systems consisting
of a huge number of relatively simple units, the set of equivalence may well be finite, see the example above. As can be seen in the example, generating
a complete system of equivalence classes for inputs might be essentially the same task as coming up with the code for the program. In fact, this
automated generation of equivalence classes should start from the problem specification and \textit{not }from a program code - as most of the commercial
``white box'' test generation programs do.
\end{itemize}

\section*{How Does Artificial Intelligence Fit into the Picture?}

Undoubtedly, in the past two decades, artificial intelligence has gained enormous attention. This is due to the fact that, by the drastically increased
computational power of current computer systems and the availability of huge databases of ``labeled'' data, a couple of difficult and urgent
problems have received impressive solutions by artificial intelligence methods, as for example machine translation of natural languages.

Amazingly, there is still a lot of mystery, nebulosity, and misunderstanding around what artificial intelligence (AI) actually is and why it is /
may be / is believed to be essentially different from all computational approaches so far. This nebulosity is all over the place: in politics, in the
media, even in science, and, of course, with the man on the street. At times, I have the impression that even quite some researchers in the AI area
do not have a very clear picture of the distinctive characteristics of AI when compared with other computational approaches. Also, labeling a project
with AI, may have a beneficial effect when it comes to funding, societal respect, political influence etc. Thus, it is tempting to keep the notion
ambiguous. What amazes me, even more, is that the nebulosity about the essence of AI did not disappear since the field started in the middle of the
fifties. I remember talks of AI evangelists around 1980, i.e. in the ``first wave of AI research'', who believed and spread that ``AI can solve
hard problems that cannot be solved by mathematics''. And still, when I participate in political discussions about the importance of mathematical
education (in the sense of training mathematical thinking), I hear the argument that, actually, the ability to do mathematics will be less and less
important because  ``tedious'' mathematical thinking, in the presence of ``artificial intelligences'' (plural!), will not be necessary
anymore and that we should teach the youngsters more ``creative'' things than mathematics. 

Now all such statements may be true or false according to which notion of artificial intelligence one has in mind. For clarifying this notion, I
want to distinguish three possible characterizations of AI:

\begin{description}
\item[\rm\textit{Hard Problems:}] 
Artificial Intelligence may be described as the field that tries to solve problems that, at a certain historic moment, are considered to be ``hard''
in the sense that they apparently need a decent amount of (human) ``intelligence'' to solve them. For example, playing chess or finding symbolic
integrals, at some historic moment, were considered as needing human intelligence. Algorithms (invented by humans!) that finally were able to solve
these problems were then (and still are) considered to be the result of ``AI research''. 
\end{description}

Now, in my opinion, this definition of the notion of AI is quite shallow. It is the natural flow of science and technology that we can solve harder
and harder problems automatically, i.e. by algorithms. However, from some point on, people think that now ``algorithms are taking over'', ``artificial
intelligence is replacing humans'' etc. forgetting that this happened and happens already since centuries and that this is the very goal of science
and technology. And, of course, whatever the methods behind automation were and are, we humans should stay in control and decide how far we let problems
be solved and decisions be taken by algorithms. Anyway, the notion of a ``hard'' problem is relative and ``hard'' problems for which an algorithmic
solution was finally found very soon are considered to be ``easy'' by the consumer. For example, car drivers nowadays take  the functionality
of a navigation system for granted. Some thirty years ago, the current functionality of navigation systems would have been considered unbelievably
intelligent. In fact, the stack of scientific findings and algorithmic techniques involved in a navigation system for guiding a driver from $A$ to
$B$ is quite deep.

In my opinion, one should not use the notion of ``artificial intelligence'' for ``finding algorithms for hard problems'' but rather continue
to call this just ``mathematical, algorithmic solution of hard problems''. Attaching the label ``AI'' to algorithms depending on whether
they solve hard or easy problems is more a question of marketing rather than a logically sound distinction.

\begin{description}
\item[\rm \textit{Simulate the Brain:}]
A completely different view (and branch) of artificial intelligence is artificial intelligence as the science of understanding and simulating biological
structures that show ``intelligence'', notably the human brain. This type of AI research, historically, was one of the origins of the field of
AI that started, maybe, 1943 with the investigations of W. McCulloch and W. Pitts who introduced a simple mathematical model of the functionality
of a neuron. Of course, understanding and simulating the most complex biological systems, which are commonly considered to display ``intelligence'',
 is a highly fascinating and relevant undertaking. Well, why not call this type of research ``artificial intelligence'' in the same way as
a technical realization of the phenomenon of flying could be called ``artificial flying''.
\end{description} 

``Artificial intelligence'' in the sense of brain simulation has little overlap with symbolic computation in software science except that, of
course, there may be applications of symbolic computation in developing models of the brain. Also, studying biological structures (like the brain,
like swarms of animals, or like the evolution of life on earth) motivated some of the algorithmic methods that today are called ``AI methods'',
see next paragraphs.

\begin{description}
\item[\rm \textit{``Intelligent'' Methods:}]
The third approach of characterizing artificial intelligence is by specifying certain algorithmic methods as ``intelligent''. These algorithms
would constitute the area of ``artificial intelligence''. I hope I do not overlook something important but my impression is that, essentially,
``machine learning'' is the only such method or, better, class of methods that has not already been around before the term ``artificial intelligence''
was coined. The many other algorithmic methods that are often labeled as ``AI methods'', like automated reasoning, semantic networks, graph search,
expert systems, regression, etc., in my view, are algorithmic methods that are not specific to AI. They are, so to say, usual algorithmic methods
and were applied also to problems that, for some reason, got the label ``AI'', see the remarks about hard problems above.
\end{description}

In my view, machine learning methods cannot actually be specified by the way how they work but, rather, by the way the problems these methods should
solve are specified. As we have seen in the previous section, the fundamental part of algorithm and software development is the transition from a
given problem specification $P$ to an algorithm (program, system) $A$ that solves the problem for any admissible input. As long
as the steps for going from a problem specification to a solution algorithm are done by a human this is just the ``usual'' business of mathematics/informatics.
If finding these steps is (partially) supported by algorithms (invented by humans) this is what we can call ``symbolic computation in software
science''. How and when does ``machine learning'' come in and why, if at all, is this different from ``usual'' mathematics and ``usual''
(maybe quite sophisticated) symbolic computation in software science? 

The point is that, in many situations, when we want to specify a problem, we do not have a specification ``in general terms'' available. For
example, let{'}s consider the seemingly simple problem of deciding whether a given English sentence contains information of the type ``somebody
cooperates with somebody else''.  An algorithm for this problem should produce the answer ``\textsc{no}'' in case no such information is in the input
sentence and should produce ``\textsc{yes}'' and the two ``somebodies'' if such information is in the sentence. Now, of course, before trying to invent
such an algorithm, we will ask: What exactly do you mean by ``cooperate''? Among the English speaking community, under the natural assumption
of a long experience of using English in thousands of different situations, it would be natural the start to explain ``cooperate'' in terms of
a couple of other notions like ``working together'', ``having a common goal'', \ldots Oh, ``having a common goal'' may not always be sufficient
for speaking about ``cooperation''. One may have a common goal but fight against each other. Thus, ``supporting each other'' etc. should
be added. Some more subtle details should be explained, some other things excluded etc. A long list of sentences explaining the meaning of  ``cooperate''
would be necessary. Then one could, in the attempt of finding an algorithm for this little problem, try to put these numerous explanations into algorithmic
rules (assuming that we already have access to a powerful grammar parsing algorithm for all of English). As a result, we would hope that this rule
system would be able to do the job. For example, if we now would input ``Peter and Ann found a way to help each other for passing the exam'',
the algorithm should answer ``\textsc{yes}'', ``Peter'', ``Ann''. If we would input ``Peter and Ann passed the exam on the same day'', it
should answer ``\textsc{no}''. Should it really answer ``\textsc{no}''? Shouldn't it rather answer ``\textsc{don't know}'' or ``\textsc{could be}'' or ``\textsc{could
be but not explicitly mentioned}''.

I now want to explain what, in my view, is the essence of the machine learning approach. For this, we need not at all bother about what ``learning''
is. I just consider those methods that, over the years, have been named ``machine learning'' methods. The common feature of these methods is
not how they proceed but the type of specification of the problems to which they are applied: Namely, they all are applied to problems of the kind
above where a spelled-out complete specification is not possible or, at least, not feasible. Now, the fundamental idea of machine learning for solving
such problems is:

\begin{itemize}
\item Instead of spending time trying to specify the problem by a huge number of general definitions, cases, rules, etc., one spends the time giving a
huge number of examples of input instances together with the answers. (In this paper, we consider only ``supervised learning''.) In this context,
the answers are called ``labels''. 

\item One sets up an algorithm from a certain class of relatively simply structured algorithms (like the class of neural networks, the class of hyperplanes
in a high-dimensional space, the class of nested if-then-else expressions, etc.) with some constants $ c_1,\ldots ,c_n$  (for example the weights
at the inputs of neurons in neural networks) in the algorithm left unspecified. For each choice of numerical values for the $ c_1,\ldots,c_n$, the algorithm would produce an answer for each admissible input for the problem.

\item One uses techniques of mathematical optimization (or other, experimental techniques, for example techniques that mimic biological evolution) to change
the initial values for  $c_1,\ldots,c_n$  iteratively until the answer of the algorithm to more and more inputs from the set of labeled data
would give the answer specified by the label. In the jargon of machine learning, this iteration is called ``training a model''. 

\item One stops the iteration on the  $c_1,\ldots ,c_n$   when sufficiently many answers are identical to the labels. Practically, at the beginning
of the whole operation, one partitions the set of labeled input into a ``training set'' that is used for the iterations and a ``test set''
on which the algorithm with the current values for the  $c_1,\ldots,c_n$  - which in the jargon of machine learning is called the ``trained
model'' - is tested.
\end{itemize}

The impressive success of this approach in the past two decades hinges on three ingredients:

\begin{itemize}
\item a huge amount of \textit{mathematical} research on good and, partly provably convergent, techniques for improving the algorithm parameters $ c_1,\ldots ,c_n$;
such research was partly already available in the first phase of AI between 1960 and 1980, but it did not convince because of the next two ingredients
were not available,

\item huge corpora of labeled data; for example, in the spectacular application of machine translation, a huge amount of ``labeled data'' is now available
in the form of files that contain an original book and its translation - by humans -  to some other language,

\item high-performance computing; the number of iterations of the machine learning steps for determining suitable   $ c_1,\ldots ,c_n$  and the computational
effort in each step is huge and is only manageable by computers in recent years.
\end{itemize}

In principle, the approach is not radically new. Examples of historical ``learning from examples'' problems are:  Given points in the plane,
 find the coefficients $ c_1,\ldots ,c_n$  of a polynomial that goes through the given points (the interpolation problem). Given a function
with some properties on differentiability, an interval, and a distance, find the coefficients $ c_1,\ldots ,c_n$  of a polynomial that is closer
to the function than the given distance everywhere on the interval (approximation problem). Given points in the plane, find the coefficients of a
straight line such that the distance to all points is minimal (regression problem). Given a function with certain differentiability properties, find
the coefficients  $ c_1,\ldots ,c_n$  of  a finite Fourier approximates of the function. Etc.

Artificial Intelligence in the form of machine learning falls neatly into the ``automatic programming'' view: It is the method of choice in cases
where the problem is not specified by general (formal or natural language) statements but, rather, is specified only by a (huge) number of examples
of admissible input and desired output. In the case of general specifications of problems, the transition from the problem to a solving algorithm,
in principle, is a reasoning process that is executed by humans or, in the symbolic computation approach advocated in this paper, is a reasoning
process (partly) executed by symbolic computation methods. In the case of problems that are specified only by examples, this process can still be
automated by the machine learning approach.

From the simple summary of the machine learning approach I gave above, one important deficiency of the machine learning approach should be clear:
The algorithm which we get for a given problem just does the job of delivering (in sufficiently many cases) desired answers. However, in general,
no reason can be given why, for example,  the particular neural network that translates one natural language to the other mimics certain fundamental
insights about the environment both languages share as their semantics. This is, in fact, similar to the situation in the historical predecessors
of ``learning from examples'': The Fourier analysis just does the job of finding an optimal Fourier sum. In the example, where the function to
be represented is the frequency spectrum of a musical tone, the representation by a finite Fourier sum has a reasonable ``explanatory'' power:
The tone is composed of tones and overtones that occur in the physical ``music'' world (for example, when picking the strings of a guitar). However,
if a Fourier representation of some arbitrary other function is executed, there will be some outcome but there may not be any reasonable interpretation
of what this representation means in the reality from which the function is taken.

The problem of weak explanatory power in the models (algorithms) created in machine learning is well known. Lots of research was recently started
to extract ``meaning'' from such models. This research area is called ``explainable AI''.

In the frame of the analysis of this paper, I summarize: The machine learning approach can be well subsumed under the general target of (semi-)automating
the process of software development (``automatic programming''). It can be viewed as a numerical, rather than a symbolic, approach to automatic
programming. Thus, it is probably a very good idea to integrate machine learning into the scope of the SCSS series because, of course, the interaction
of symbolic and numerical computation as the two possible approaches to compute on finitary representations of abstract mathematical domains is of
utmost importance.  The integration of machine learning into the scope of SCSS can generate a stream of new ideas in both directions: Applying
symbolic methods to mathematical sub-problems of machine learning (e.g. the determination of weights in neural networks) and applying machine learning
to symbolic algorithms (e.g. ``learning'' a priori complexity estimates for computation-intensive methods like Gr{\"o}bner bases etc.).

However, there is no reason to establish a flavor of ``intelligence beyond mathematics'' when speaking about machine learning: I hope I have
been able to show the machine learning is just another mathematical method.  As in the past, of course, we can hope and expect also for the future
that more and more powerful algorithmic problem-solving methods will be invented. 

Personally, when speaking to people who do not (want to) understand the timeless, universal, always new, creative power of mathematics, I like to
use the term \textit{algorithmic intelligence }for what we are doing: Algorithmic intelligence is the \textit{human} intelligence that produces
algorithms for more and more challenging problems in all areas of human activity. By an algorithm, an infinite class of individual problem instances
can then be treated by a completely \textit{unintelligent} machine. People who do not really understand what is going on may \textit{believe that
	these machines display ``intelligence''}. The algorithmic intelligence - by reflection, i.e. jumps to higher and higher meta-levels - also provides
more and more sophisticated algorithms for producing algorithms.  The incompleteness theorem of Kurt G{\"o}del (1931), in a somewhat liberal interpretation,
shows that this tour through higher and higher levels of algorithmization has no upper bound. In comparison to ``artificial intelligence'', the
term ``algorithmic intelligence'' is used quite rarely, which can be verified by Googling. However, my impression is that ``algorithmic intelligence''
appears in more serious discussions about the essence of AI. Therefore, I like to expand the abbreviation ``AI'' as ``algorithmic intelligence''.

Implicitly, I expressed this view already in the early days of AI, see~\cite{DBLP:conf/ki/Buchberger82}. At the ``Spring School of AI'' in Teisendorf, 1982, I contributed
a long article summarizing the most important ``symbolic'' methods for automating the algorithm/software development process that were known
at that time. And we had long, intensive, and quite controversial discussions at this conference on the question of whether AI is something that
goes beyond mathematics. As you may guess, my answer then was ``no'' with essentially the arguments which I expanded above. In my hectic years
of research on methods for ``algorithmic intelligence'' and research management, I never found the time and occasion to spell out these arguments
in a paper. Thus, I am grateful that I am given the opportunity here.


\begin{thebibliography}{1}
	\providecommand{\bibitemdeclare}[2]{}
	\providecommand{\surnamestart}{}
	\providecommand{\surnameend}{}
	\providecommand{\urlprefix}{Available at }
	\providecommand{\url}[1]{\texttt{#1}}
	\providecommand{\href}[2]{\texttt{#2}}
	\providecommand{\urlalt}[2]{\href{#1}{#2}}
	\providecommand{\doi}[1]{doi:\urlalt{http://dx.doi.org/#1}{#1}}
	\providecommand{\bibinfo}[2]{#2}
	
	\bibitemdeclare{inproceedings}{DBLP:conf/ki/Buchberger82}
	\bibitem{DBLP:conf/ki/Buchberger82}
	\bibinfo{author}{Bruno \surnamestart Buchberger\surnameend}
	(\bibinfo{year}{1982}): \emph{\bibinfo{title}{Computer-unterst{\"{u}}tzter
			Algorithmenentwurf (Computer-Aided Algorithm Design)}}.
	\newblock In \bibinfo{editor}{Wolfgang \surnamestart Bibel\surnameend} \&
	\bibinfo{editor}{J{\"{o}}rg~H. \surnamestart Siekmann\surnameend}, editors:
	{\sl \bibinfo{booktitle}{Proceedings of the Fr{\"{u}}hjahrsschule
			K{\"{u}}nstliche Intelligenz (Spring School in Artificial Intelligence),
			Teisendorf, Germany, 15.-24. M{\"{a}}rz 1982}}, {\sl
		\bibinfo{series}{Informatik-Fachberichte}}~\bibinfo{volume}{59},
	\bibinfo{publisher}{Springer}, pp. \bibinfo{pages}{141--201},
	\doi{10.1007/978-3-642-68828-7\_4}.
	
	\bibitemdeclare{article}{jsc1985editorial}
	\bibitem{jsc1985editorial}
	\bibinfo{author}{Bruno \surnamestart Buchberger\surnameend}
	(\bibinfo{year}{1985}): \emph{\bibinfo{title}{Symbolic Computation (An
			Editorial)}}.
	\newblock {\sl \bibinfo{journal}{Journal of Symbolic Computation}}
	\bibinfo{volume}{1}(\bibinfo{number}{1}), pp. \bibinfo{pages}{1--6},
	\doi{10.1016/S0747-7171(85)80025-0}.
	
	\bibitemdeclare{inproceedings}{scss2013}
	\bibitem{scss2013}
	\bibinfo{author}{Bruno \surnamestart Buchberger\surnameend}
	(\bibinfo{year}{2013}): \emph{\bibinfo{title}{Mathematics of the 21st
			Century: A Personal View}}.
	\newblock In \bibinfo{editor}{Laura \surnamestart Kov\'{a}cs\surnameend} \&
	\bibinfo{editor}{Temur \surnamestart Kutsia\surnameend}, editors: {\sl
		\bibinfo{booktitle}{Proceedings of the Fifth International Symposium on
			Symbolic Computation in Software Science ({SCSS 2013)}}}, {\sl
		\bibinfo{series}{RISC Report Series}} \bibinfo{volume}{13-06},
	\bibinfo{organization}{Johannes Kepler University},
	\bibinfo{address}{Linz/Hagenberg, Austria}, p.~\bibinfo{pages}{1}.
	\newblock
	\urlprefix\url{https://www.risc.jku.at/publications/download/risc_4737/TR_13_06_SCSS2013.pdf}.
	\newblock \bibinfo{note}{(See also the link to the slides of this talk on the
		website of SCSS 2013 at
		\url{https://www.risc.jku.at/conferences/scss2013/program.html}.)}.
	
	\bibitemdeclare{proceedings}{mkm01}
	\bibitem{mkm01}
	\bibinfo{editor}{Bruno \surnamestart Buchberger\surnameend} \&
	\bibinfo{editor}{Olga \surnamestart Caprotti\surnameend}, editors
	(\bibinfo{year}{2001}): \emph{\bibinfo{title}{Electronic Proceedings of the
			First International Workshop on Mathematical Knowledge Management ({MKM
				2001})}}. \bibinfo{organization}{RISC, Johannes Kepler University},
	\bibinfo{address}{Linz/Hagenberg, Austria}.
	\newblock
	\urlprefix\url{https://www.risc.jku.at/conferences/MKM2001/Proceedings/}.
	
	\bibitemdeclare{article}{DBLP:journals/amai/BuchbergerGH03}
	\bibitem{DBLP:journals/amai/BuchbergerGH03}
	\bibinfo{author}{Bruno \surnamestart Buchberger\surnameend},
	\bibinfo{author}{Gaston~H. \surnamestart Gonnet\surnameend} \&
	\bibinfo{author}{Michiel \surnamestart Hazewinkel\surnameend}
	(\bibinfo{year}{2003}): \emph{\bibinfo{title}{Mathematical Knowledge
			Management.}}
	\newblock {\sl \bibinfo{journal}{Special issue of the Annals of Mathematics and
			Artificial Intelligence}} \bibinfo{volume}{38}(\bibinfo{number}{1-3}), pp.
	\bibinfo{pages}{1--2}, \doi{10.1023/A:1022900528196}.
	
\end{thebibliography}

\end{document}